\documentclass[conference]{IEEEtran}
\IEEEoverridecommandlockouts

\usepackage{amsmath,amsfonts,amssymb}
\usepackage{algorithmic}
\usepackage{array}
\usepackage{booktabs}
\usepackage[caption=false]{subfig}
\usepackage{textcomp}
\usepackage{url}
\usepackage{verbatim}
\usepackage{graphicx}
\usepackage{soul}
\usepackage{xcolor}
\usepackage{siunitx}
\usepackage{balance}

\usepackage{tikz}
\usetikzlibrary{shapes, arrows, positioning, backgrounds, calc, matrix}

\usepackage{adjustbox}
\usepackage{bm}

\usepackage{pgfplots}
\pgfplotsset{compat=1.18}

\DeclareMathOperator*{\argmax}{arg\,max}

\usepackage{preamble}
\usepackage[acronym,style=alttree,nomain,shortcuts,nonumberlist,nogroupskip]{glossaries}
\newacronym{3d}{3D}{three-dimensional}
\newacronym{6G}{6G}{sixth generation}

\newacronym{ae}{AE}{autoencoder}
\newacronym{aoa}{AoA}{angle of arrival}
\newacronym{ap}{AP}{access point}

\newacronym{bs}{BS}{base station}

\newacronym{ckm}{CKM}{channel knowledge map}
\newacronym{cir}{CIR}{channel impulse response}
\newacronym{csi}{CSI}{channel state information}

\newacronym{glrt}{GLRT}{generalized likelihood ratio test}
\newacronym{gnn}{GNN}{generative neural network}

\newacronym{ekf}{EKF}{extended Kalman filter}

\newacronym{kf}{KF}{Kalman filter}

\newacronym{isac}{ISAC}{integrated sensing and communication}

\newacronym{los}{LoS}{line-of-sight}
\newacronym{lrt}{LRT}{likelihood ratio test}
\newacronym{ls}{LS}{least squares}

\newacronym{mimo}{MIMO}{multiple input multiple output}
\newacronym{ml}{ML}{maximum likelihood}
\newacronym{music}{MUSIC}{multiple signal classification}

\newacronym{ocsvm}{OC-SVM}{one-class support vector machines}

\newacronym{pl}{PL}{path loss}
\newacronym{pla}{PLA}{physical layer authentication}

\newacronym{rnn}{RNN}{recurrent neural network}

\newacronym{snr}{SNR}{signal to noise ratio}

\newacronym{uav}{UAV}{unmanned aerial vehicle}
\newacronym{ue}{UE}{user equipment}
\newacronym{uwb}{UWB}{ultra-wide band}

\newacronym{v2x}{V2X}{vehicle-to-everything}

\newacronym{rrc}{RRC}{root raise cosine}

\newacronym{ula}{ULA}{uniform linear array}
\newacronym{ura}{URA}{uniform rectangular array}
\newacronym{awgn}{AWGN}{additive white Gaussian noise}

\newacronym{det}{DET}{detection error trade-off}
\newacronym{auc}{AUC}{area under the curve}
\newacronym{spla}{SPLA}{static physical layer authentication}
\newacronym{kpla-los}{KPLA-Los}{Kalman physical layer authentication - line of sight}
\newacronym{kpla-ckm}{KPLA-CKM}{Kalman physical layer authentication - channel knowledge maps}
\newacronym{rmse}{RMSE}{root mean square error}
\newacronym{lm}{LM}{link margin}
\newacronym{pdf}{PDF}{probability distribution function}
\newacronym{urp}{URP}{useful received power}
\newacronym{lt}{LT}{likelihood test}
\newacronym{ra}{RA}{range-angle}
\newacronym{toa}{TOA}{time of arrival}

\begin{document}

\title{ISAC-Assisted Channel Knowledge Map Generation for Physical Layer Authentication}

\author{ \IEEEauthorblockA{	
Luca~Bonaventura*, Edoardo~Gardin, Alessia~Barison, Francesco~Ardizzon, and Stefano~Tomasin\\ 
}\vspace{2mm}
 \IEEEauthorblockA{Department of Information Engineering, University of Padova, Italy\\
 \small $^\star$ Corresponding author, email: luca.bonaventura@phd.unipd.it
 }\thanks{This work was supported by Agenzia per la cybersicurezza nazionale under the programme for promotion of XL cycle PhD research in cybersecurity - C96E24000010005. The views expressed are those of the authors and do not represent the funding institution. 
}}

\maketitle

\begin{abstract}
\Ac{isac} enables the acquisition of environmental information by leveraging wireless signals transmitted for communication purposes. In this paper, we utilize this capability to reconstruct the layout of objects surrounding multiple receivers. Ray tracing is then applied to the reconstructed environment to infer the propagation channels for various transmitter positions, thereby constructing a \ac{ckm}. The \ac{ckm} is then used to verify the position of a legitimate transmitter, authenticating it against an adversarial device attempting to impersonate it from a different location. This \ac{pla} mechanism utilizes the approximate known position of the legitimate transmitter, obtained, for instance, from the network as in cross-layer authentication, to compare the channel estimated from the received signal with the corresponding \ac{ckm} data. We evaluate the impact on the \ac{pla} performance of both \ac{isac}-induced \ac{ckm} reconstruction errors and receiver-side channel estimation noise, in terms of false alarm and missed detection probabilities. Finally, the proposed approach is validated using an \ac{isac} dataset from the literature. 

\end{abstract}


\glsresetall

\section{Introduction}\label{sec:intro}

\Ac{isac} is expected to provide new services in \ac{6G} networks \cite{10979960}, and it can be used to reconstruct the propagation environment in which communication occurs, \cite{10430216}. In this paper, we exploit \ac{isac} to obtain a novel \ac{pla} mechanism that enables the \ac{bs} of a \ac{6G} network to authenticate received signals, i.e., to confirm that they come from a specific legitimate source. To this end, we exploit \ac{isac} to first reconstruct the environment and then to obtain a \ac{ckm}~\cite{Esrafilian2019Learning}  that enables the network to know in advance the channel expected for \acp{ue} in any position. By verifying real-time channel observations against those obtained from the \ac{ckm}, the resulting \ac{isac}-based \ac{pla} mechanism provides lightweight security with zero communication overhead, making it uniquely ideal for energy-constrained or rate-limited devices. However, establishing this required a priori model remains highly challenging for mobile users due to rapid temporal variations, as well as in complex indoor settings where non-\ac{los} components and dynamic physical obstructions predominate.

While \acp{ckm} have been applied to optimize \ac{uav} networks~\cite{Haoyun2023Channel} and enable training-less beamforming~\cite{Dai2024Prototyping}, and can even be constructed dynamically using \ac{isac}~\cite{ZhangChaoyue-2025}, their exploitation for physical layer security remains largely unexplored. The few existing \ac{ckm}-based authentication strategies, e.g., in~\cite{Wang2024Knowledge}, rely on the highly restrictive assumption that the legitimate user follows a trajectory predefined a priori by the verifier, limiting practical, uncoordinated deployments. In~\cite{Bonaventura2026PLA}, we designed a \ac{pla} mechanism for authentication with \ac{ckm}. However, we did not account for the errors and challenges of an actual ISAC-derived \ac{ckm}. In this work, we design a framework that also addresses these challenges. Among other aspects, we consider the sensing-based map derivation and PLA to occur at different carrier frequencies and thus may be affected by different noises. Furthermore, we consider that part of the environment may not be reconstructed correctly due to obstructions or the shape of the target. This requires a \ac{ckm} generation pipeline passing through point cloud reconstruction, \ac{3d} environment reconstruction, and ray tracing.

To overcome these limitations, this paper introduces a novel, environment-aware \ac{pla} framework leveraging \ac{isac}-driven \acp{ckm} without requiring predefined user trajectories, but only partial information, e.g. having a coarse position from the upper layers in cross-layer authentication, or knowing the previous position of the user, as in \cite{Bonaventura2026PLA}. The key contributions of this work are summarized as follows:
\paragraph*{\ac{isac}-Based Environment Reconstruction} We utilize \ac{isac} signals across multiple receivers to dynamically reconstruct the layout of surrounding objects, mapping the localized environment without dedicated sensing infrastructure using Poisson~\cite{Kazhdan-2006} with spatial tapering~\cite{Liu-2025}.
\paragraph*{Ray Tracing-Based \ac{ckm} Generation} We apply a deterministic ray tracing engine to the reconstructed object layout to predict site-specific propagation channels across various locations, establishing a robust, location-dependent \ac{ckm}.
\paragraph*{Position-Based \ac{pla} and Refinement} We design a verification mechanism that authenticates a legitimate transmitter and detects spoofing attacks from alternative locations by comparing estimated channels with \ac{ckm} entries, subsequently refining the authenticated transmitter's position estimate.
\paragraph*{Performance Evaluation under Impairments} We evaluate, through Monte Carlo simulations, the impact of \ac{isac} reconstruction errors and receiver-side channel estimation noise on false alarm and missed detection probabilities, validating the framework using a literature-sourced \ac{csi} database.

The remainder of the paper is organized as follows. Section~\ref{sec:sysModel} details the system model. Section~\ref{sec:ISACCKM} describes the \acp{ckm} derivation via \ac{isac}. The considered \ac{pla} mechanism is introduced in Section~\ref{sec:PropScheme}. Section~\ref{sec:NumResults} presents the numerical results. Section~\ref{sec:Concl} draws the conclusion.

\section{System Model}\label{sec:sysModel}

We consider a cellular network comprising $N$ \acp{bs} performing \ac{isac} transmissions to simultaneously communicate with the \acp{ue} and sense the surrounding environment for the construction of a \ac{ckm}, as described in the following. A monostatic \ac{isac} architecture is assumed, in which each \ac{bs} is equipped with two distinct \acp{ura}: one transmit \ac{ura} with $N_{\rm t} \times N_{\rm t}$ antenna elements and one receive \ac{ura} with $N_{\rm r} \times N_{\rm r}$ antenna elements. The \acp{bs} are assumed to be loosely synchronized, and their connections to the core network are considered secure. 

\paragraph*{Channel Characteristics} To capture the essence of our \ac{pla} mechanism based on \acp{ckm}, we consider a static environment, where the surrounding scene is assumed to remain stationary during the observation interval. Thus, the propagation characteristics are time-invariant. Transmissions are narrowband. The symbol period is denoted as $T_\mathrm{s}$.

Each \ac{bs} periodically transmits a known probing waveform, with the delay applied across its \ac{ura} as a function of the azimuth $\theta$ and elevation $\phi$ angles, respectively, denoted by $s(t, \theta, \phi)$. The signal $s$ has carrier frequency $f_{\rm I}$ and bandwidth $B_{\rm I}$. The signal propagates through the environment and is subsequently reflected by surrounding scatterers. The receiver is assumed to be affected by \ac{awgn} noise with variance $\sigma^2_{M}$. The received signal is
\begin{equation}
    r(t, \theta, \phi) = \tilde{h}(t) \ast s(t, \theta, \phi) +w(t)\,,
\end{equation}
where $\ast$ denotes convolution, $\tilde{h}(t)$ is the \ac{cir} of the reflection channel and $w(t)$ is complex Gaussian distributed with zero mean and variance $\sigma^2_{\rm M}$, i.e., $w(t)\sim\mathcal{CN} (0, \sigma^2_{\rm M})$.


\begin{figure}
    \centering
    \begin{adjustbox}{width=0.5\textwidth, keepaspectratio}
    \input{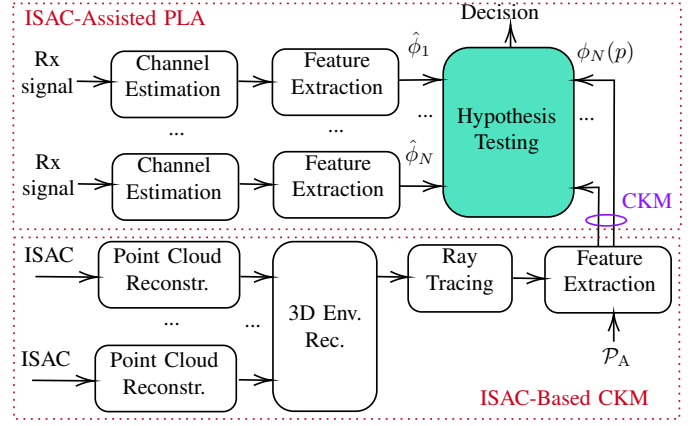}
    \end{adjustbox}%
    \caption{Schematic representation of the proposed \ac{isac}-based \ac{pla} mechanism.}
    \label{fig:CSI_scheme}
    \vspace{-.5cm}
\end{figure}
\subsection{Security Scenario}

For the design and evaluation of our \ac{pla} mechanism, we consider that a legitimate \ac{ue}, Alice, is moving in the environment, communicating with the \acp{bs}. An attacker, Trudy, is instead a \ac{ue} transmitting messages that aim at impersonating (or spoofing) Alice. On the basis of the signals received by all the \acp{bs}, the network will decide if the received signals come from Alice or Trudy. 

The \ac{ue} positions are represented on a two-dimensional horizontal plane describing the environment. The plane is discretized into square cells of side length $W_M$. Each position $\bm{p}=(x,y)$ corresponds to the center of one cell, and the set of all admissible positions is denoted by $\mathcal{P}$.  

\paragraph*{Assumptions on Alice and Bob}  We assume Bob to have partial information about Alice, e.g.,  by knowing Alice's previous position or by receiving such information from the upper layer in a cross-layer authentication setting. Thus, while Alice's true position $\bm{p}_{\rm{A}}$, is unknown to both Bob and Trudy, Bob knows that Alice position lies in the region $\mathcal{P}_{\rm A} = \{\bm{p}\in\mathcal{P}| \|\bm{p} - \bm{p}_{\rm A} \| \leq W_{\rm A}\}$, where $W_{\rm A}$ represents the accuracy on the Alice's reported position. 

Bob shall then combine such knowledge about Alice's position with the \acp{ckm} at all the \acp{bs} to test whether the received signals are authentic or not. Authentication in particular is performed on the estimated channel from Alice's transmissions, obtained through publicly known $N_{\rm p}$ pilot sequences, exploited by Bob for channel estimation. It is worth remarking that Alice’s transmission occurs at carrier frequency $f_{\rm A}$, while the map derivation occurs via sensing at carrier frequency $f_{\rm I} \geq f_{\rm A}$.

\paragraph*{Assumptions on Trudy} 
We assume that Trudy has full knowledge of the authentication protocol, including signal format and pilot symbols. 

\section{ISAC-Based Channel Knowledge \\ Map Estimation}\label{sec:ISACCKM}

The scheme of the proposed \ac{isac}-based \ac{pla} mechanism is shown in Fig.~\ref{fig:CSI_scheme}. It comprises two main parts: the lower part performs the \ac{ckm} estimation, while the upper part performs the \ac{pla} by exploiting the \ac{ckm} and the \ac{csi} estimated from the received signal to be authenticated. This section describes the operations performed in the lower part to obtain the \ac{ckm} from the environment reconstructed from the \ac{isac} signals. Section~\ref{sec:PropScheme} will instead detail the \ac{pla} part.

The \ac{ckm} estimation comprises several blocks: first, the \ac{isac} signal is processed to obtain a point cloud of the obstacle present in the environment. Then, a \ac{3d} reconstruction of the environment is obtained by using Poisson reconstruction. Such reconstruction is used to predict the channel conditions in the position of Alice, using a ray-tracer, thus providing the \ac{ckm}. In the following, we describe the various blocks in detail.

\subsection{Point Cloud Reconstruction}

To reconstruct the environment, we resort to a \ac{ra} reconstruction~\cite{Luo-2025}. Following~\cite{Liu-2025}, we first compute \ac{toa} and the intensity of the highest peak at the $n$-th \ac{bs} as 
\begin{align}\label{eq:ToaEst}
    T_n( \theta, \phi) &= \argmax_t r_n(t,\theta, \phi) \ast s(t,\theta, \phi)\,,\\
    I_n( \theta, \phi) &=\int r_n(t,\theta, \phi) \, s^\star\left(t - T_n( \theta, \phi),\theta, \phi\right) dt \, .
\end{align}
The distance of an object from the \ac{bs} is then
\begin{equation}
d_n( \theta, \phi) = c T_n( \theta, \phi)/2 \, ,
\end{equation}
where $c$ denotes the speed of light in vacuum.
The point cloud is then defined as the set of tuples $\mathcal{C}_n = \left\{\left(d_n( \theta_i, \phi_i), \theta_i, \phi_i\right)\right\}$.

The reconstructed point cloud exhibits spurious points distributed along spherical surfaces that surround strong reflectors. These artifacts originate from the finite angular resolution of the antenna array and, more specifically, from the presence of spatial sidelobes in the beamforming response.


The reconstruction algorithm estimates the position of a scatterer by identifying, for each steering direction, the strongest peak of the channel impulse response \eqref{eq:ToaEst}. However, when a reflection is captured through a sidelobe, the estimated range remains approximately equal to that of the actual target, while the associated angular coordinates correspond to the current steering direction rather than the true one. As a result, the reconstruction becomes sidelobe-interference limited, since the same physical reflector may be repeatedly detected across adjacent steering directions. As the beam sweeps across neighboring angles, multiple points are therefore reconstructed at the same distance but with incorrect angular coordinates, producing circular arcs or spherical caps centered on the sensing \ac{bs}.

To suppress these artifacts, we use spatial tapering~\cite{Bekkerman-2006} prior to beamforming. Spatial tapering consists of weighting the antenna elements with a non-uniform amplitude distribution, thereby reducing the sidelobe levels of the array pattern. Unlike the conventional rectangular weighting, where all antennas contribute equally, tapering assigns smaller weights to elements near the array boundaries and larger weights to those near the center. This reduces the \ac{ura} sensitivity to signals arriving from undesired directions, significantly attenuating reflections captured through sidelobes~\cite{Ghosh-2025}.

In this work, a Hann window is adopted due to its favorable sidelobe suppression characteristics and smooth transition to zero at the array edges~\cite{Enggar-2016}. 
In particular, since the sensing platform employs a $N_{\rm r} \times N_{\rm r}$ \ac{ura}, we consider a 2D Hann window obtained as $\bm{W}_{\rm 2D}=\bm{w}_{\theta}\bm{w}_{\phi}^\top$, where $\bm{w}_{\theta}$ and $\bm{w}_{\phi}$ are the 1D Hann windows.


\subsection{3D Environment Reconstruction}\label{subsec:building_reconstruction}
The $N$ point clouds estimated at the individual \acp{bs} are merged into a single unified point cloud. To obtain a 3D representation from the point clouds, a mesh representation was adopted, i.e., a geometric structure composed of vertices, edges, and faces that approximates the surface of the real object. 
Poisson reconstruction~\cite{Kazhdan-2006} is used to this end, due to its robustness to noise and the capability of reconstructing surfaces without imposing constraints on their shapes.

On the other hand, one limitation of the Poisson reconstruction is the possible generation of spurious faces in regions where no points were measured, e.g., due to the presence of sparse outliers in such areas. To address this, we performed a point cloud selection procedure where we first computed the centroid of each triangular mesh. Then we compute the distance between the centroid and each vertex, and we eliminate the most distant points up to a fraction of 1\% of the overall points.
The reconstructed scenario is saved in the common STL format to aid cross-portability and visualization.

\subsection{Ray Tracing}\label{sec:rayTracing}
In the considered framework, the \ac{isac} system is employed to sense the surrounding environment and reconstruct a geometric representation of the scene, including the locations of relevant reflecting objects and scatterers. The extracted environmental information is then incorporated into a channel modeling stage based on ray tracing, implemented using the MATLAB ray tracing toolbox, to estimate the propagation channel \ac{cir} between \ac{bs} and the \ac{ue}. We denote the \ac{cir} from the \ac{ue} in position $\bm{p}$ to the $a$-th antenna of the $n$-th \ac{bs} as ${g}_{ a,n}(\bm{p})$. From our previous work~\cite{Bonaventura2026PLA} we have that ${g}_{ a,n}(\bm{p})$ is Gaussian distributed with variance $\sigma^2_{\rm M}$.

\subsection{Feature Estimation}\label{sec:featEstim}

To perform \ac{pla}, we consider two characterizing features of the \ac{csi}, namely the \ac{aoa} and the \ac{pl} of the dominant path.

For transmit (receive) antenna gains $G_{\rm T}$ ($G_{\rm R}$), and transmit power $P_{\rm T}$, the \ac{pl} estimate can then be obtained as~\cite{Sijbers-1998}
\begin{align}
\hat{r}_n(\bm{p}) &=\frac{P_{\rm T}\,  G_{\rm T}\,  G_{\rm R}}{ L_{\mathrm{D}} |\hat{\mu}_n(\bm{p})|^2},
\end{align}
where $L_{\mathrm{D}}$ represents additional attenuation factors that are not captured by the path-loss term, and 
\begin{equation}
|\hat{\mu}_n(\bm{p})|^2 = \biggr[\frac{1}{N_\mathrm{A}} \sum_{a =0}^{N_\mathrm{A}-1} |{g}_{a,n}(\bm{p})|^2 \biggr] - \sigma_{\rm M}^2 .
\end{equation}

The \ac{ml} estimate of the \ac{aoa} is obtained by scanning a discrete angular grid $\zeta \in \mathcal Z  = \{0,\ldots,\pi\}$, and computing~\cite{Stoica-1989}
\begin{equation}
\hat{\theta}_n(\bm{p}) = \argmax_{\zeta \in \mathcal Z} \Bigg|\sum_{a=0}^{N_{\rm r}^2-1} {g}_{a,n}(\bm{p}) \alpha_a(\zeta) \Bigg| ,
\end{equation}
where $\alpha_a(\zeta) =\exp\left({-j2 \pi d \sin\left(\zeta\right) a/\lambda}\right)$, $d$ is the antenna spacing and $\lambda$ is carrier wavelength.

Following the results of our previous work~\cite{Bonaventura2026PLA}, we can show that for a sufficiently large number of antennas $N_{\rm r}^2$, $\hat{r}_n(\bm{p})$ can be approximated as real Gaussian distributed, i.e.,  
\begin{equation}\label{eq:CLTPL}
\hat{r}_n(\bm{p})\sim \mathcal{N}\left(r_n(\bm{p}), \sigma^2_{\mathrm{r},n}(\bm{p})\right).
\end{equation}

Analogously, the resulting \ac{aoa} estimate at the $n$-th \ac{bs} can be approximately distributed as
\begin{align}\label{eq:CLTAoA}
\hat{\theta}_n(\bm{p})  \sim \mathcal{N}\left( \theta_n(\bm{p}), \sigma^2_{\mathrm{\theta},n}(\bm{p})\right),
\end{align}
where $\theta_n(\bm{p})$ denotes the angle of arrival associated with the strongest propagation path, i.e., the first ray.

For each \ac{bs} $n$, we define, for a given position $\bm{p}$, ${{\phi}}_{n} (\bm{p}) = \left\{\hat{r}_{n}(\bm{p}) ,\,  \hat{\theta}_{n}(\bm{p})\right\}$. The \ac{ckm} is obtained by aggregating the estimates from all the \acp{bs}, obtaing for each position $\bm{p}$ the map vector  ${\bm{\phi}}(\bm{p}) = [{\phi}_{1}(\bm{p}), \dots,{\phi}_{N}(\bm{p}) ]^\top$.

\section{\ac{isac}-based \ac{pla}  Mechanism}\label{sec:PropScheme}

The upper part of the proposed \ac{isac}-based \ac{pla} mechanism, shown in Fig.~\ref{fig:CSI_scheme}, is executed for every message received by the network. It aims at checking if the received message comes from Alice or not, and it is based on the \ac{csi} estimated at each \ac{bs}. Features relevant for \ac{pla} are extracted from the \ac{csi}, similarly to what is done to build the \ac{ckm}. Then, a hypothesis testing block compares the extracted features with those available in the \ac{ckm} and makes a decision on the authenticity of the received message. In the following, we detail the operations performed in each block.

\subsection{Channel Estimation}
Following the results of our previous work~\cite{Bonaventura2026PLA}, using the $m$-th received pilot signal, with $ m \in \{0,\dots, N_{\mathrm{p}}-1\}$, the \ac{bs} obtains an estimate $\hat{g}_{a,n}^{(m)}$ of the channel from the transmitter in position $\bm{p}$ that can be modeled as 
\begin{equation}
    \hat{g}_{a,n} = \sum_{m =0}^{N_\mathrm{p}-1}\hat{g}_{a,n}^{(m)} = h_{a,n}\big(T_{1,n}(\bm{p}), \bm{p} \big) + w,
\end{equation}
where $T_{1,n}(\bm{p})$ is the \ac{toa} of the dominant path and $w \sim \mathcal C\mathcal N(0,\sigma_{\rm C}^2)$.

\subsection{Feature Extraction}\label{sec:FeatExtr}

The \ac{bs} extracts the features using the same procedure illustrated in Section~\ref{sec:featEstim}, applied to the actual received signal $\hat{g}_{a,n}(\bm{p})$. Also in this case, the \ac{awgn} model \eqref{eq:CLTPL} holds; however, now the noise variance is $\sigma_{\rm C}^2$.

Thus, each \ac{bs} obtains \ac{pl} and \ac{aoa} estimates $\hat{\phi}_n=\{\hat{r}_n,\hat{\theta}_n\}$, that Bob combines into the global feature vector $\hat{\bm{\phi}}= [\hat{\phi}_1,\ldots,\hat{\phi}_N]^\top $.

\begin{figure}
\centering
\subfloat[Ground truth.]{\includegraphics[width=4.3cm,height=4cm,trim={0cm 0cm 0cm 1.5cm}, clip]{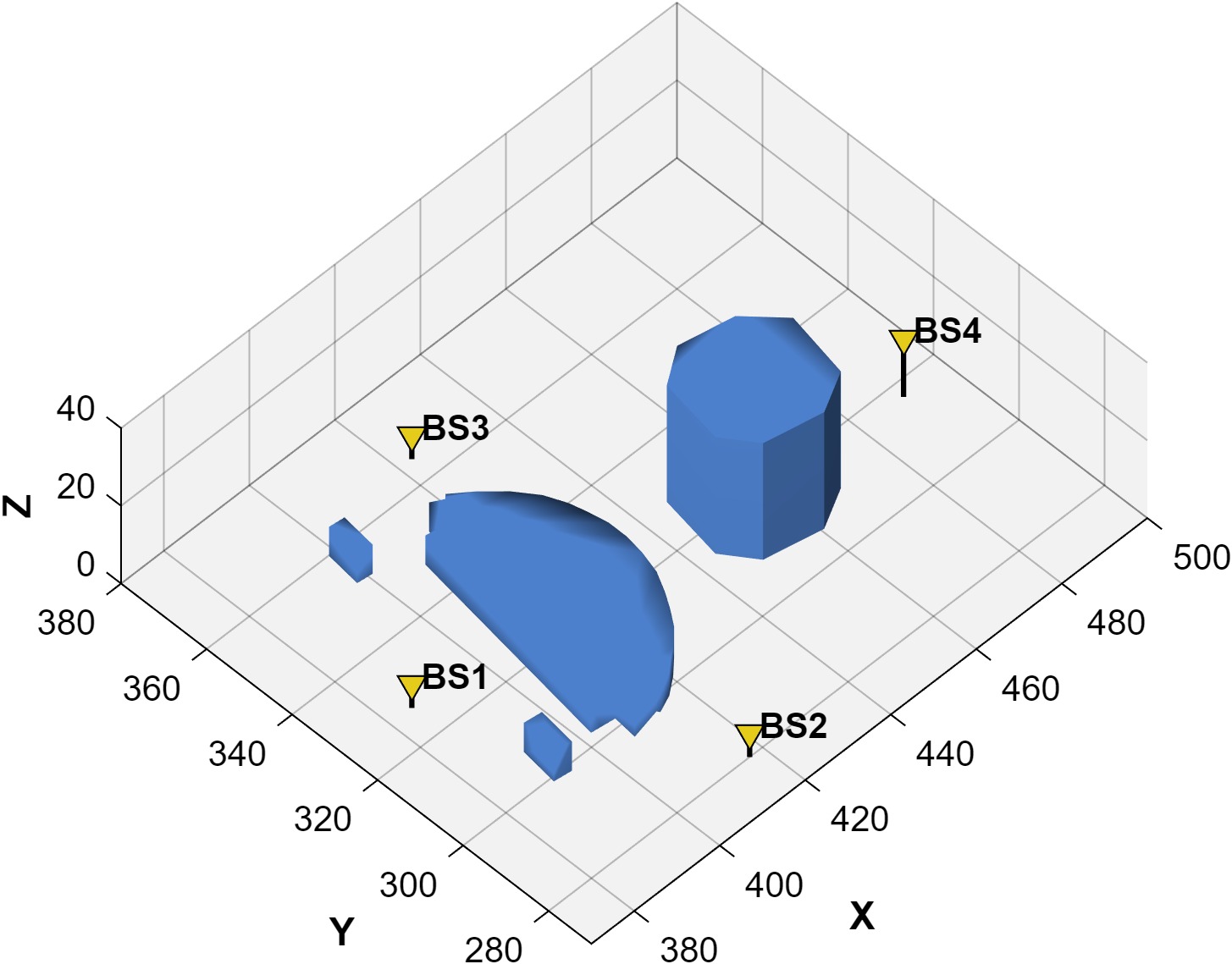}\label{fig:MapGT}}
\hfill
\subfloat[Reconstructed.]{\includegraphics[width=4.3cm,height=4cm,trim={0cm 0cm 0cm 1.5cm}, clip]{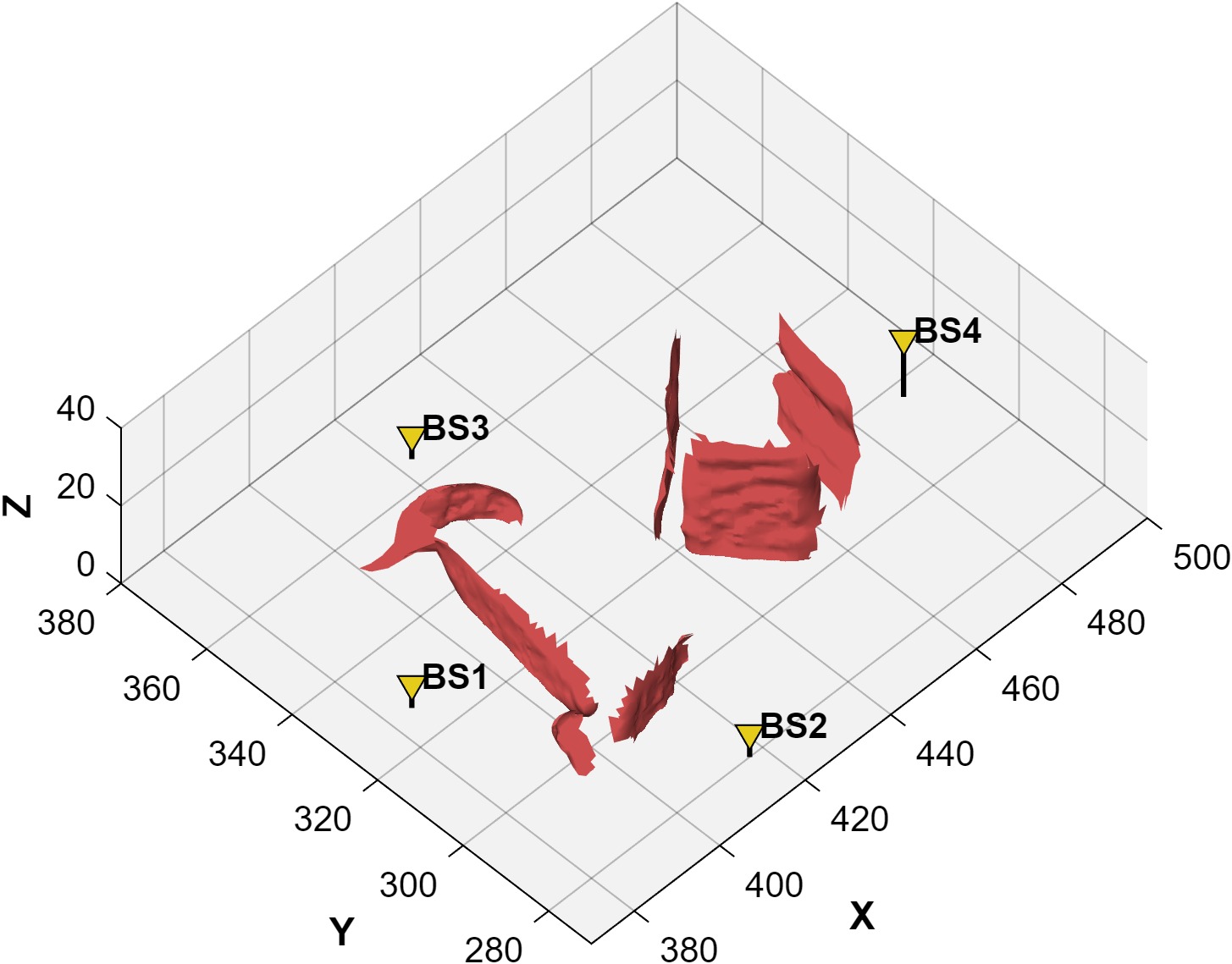}\label{fig:MapREC}}
\caption{Ground truth (a) and \ac{isac}-reconstructed environment (b), and \ac{bs} positions in the scene.}
\label{fig:ISAC_reconstructed3D}
\end{figure}
\subsection{Hypothesis Testing-based PLA Check}




The \ac{pla} uses the \acp{ckm} (Section \ref{sec:ISACCKM}) and the partial knowledge about Alice's position, $\mathcal{P}_{\rm A}$ as prior information to check whether observation $\hat{\bm{\phi}}$ (Section \ref{sec:FeatExtr}) is legitimate, thus authenticating the received signal.

The authentication test is based on binary hypothesis testing. We call $\mathcal{H}_0$ and $\mathcal{H}_1$ the legitimate and under attacker hypotheses, i.e., when Alice or Trudy is the transmitter, respectively. We assume that the legitimate party has no knowledge about Trudy. Additionally, only partial information about Alice is available. Thus, instead of the (optimal) \ac{lrt} we resort to a \ac{glrt}, where denoting with $\phi$ the noiseless channel feature vector we have
\begin{equation} \begin{split} \label{eq:likelihood}
\gamma&= \max_{\bm{p}\in\mathcal{P}_{\rm A}}  p\left( \hat{\bm{\phi}}  \Big| \bm{\phi} = \bm{\phi}(\bm{p}) \right) =
\\ &= \max_{\bm{p}\in\mathcal{P}_{\rm A}} \prod_{n=1}^N p\left(\hat{r}_n | r_n={r}_n(\bm{p})\right)   p\left(\hat{{\theta}}_n | {\theta}_n  = {\theta}_n(\bm{p})\right).
\end{split} \end{equation}
Thus, we test whether the features $\hat{\bm{\phi}}$ from the signal to authenticate match ones potentially received from a transmitter in $\mathcal{P}_{\rm A}$ and thus, legitimate.

\begin{figure*}[ht!]
\centering
\subfloat[Ground truth.]{\includegraphics[width=8cm,height=6cm,trim={3.5cm 8.5cm 3.8cm 8.7cm}, clip]{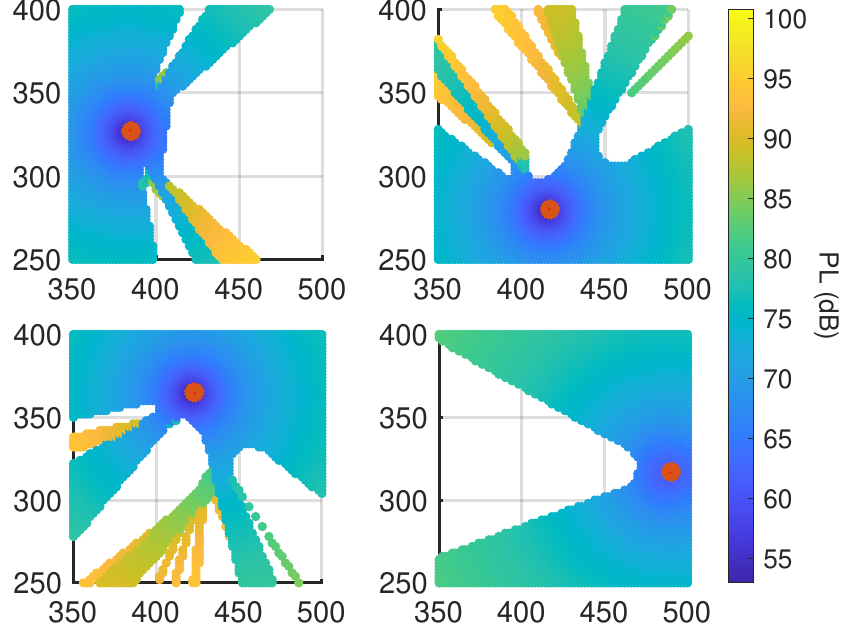}\label{fig:PLGT}}
\hfill
\subfloat[\ac{ckm}.]{\includegraphics[width=8cm,height=6cm,trim={3.5cm 8.5cm 3.8cm 8.7cm}, clip]{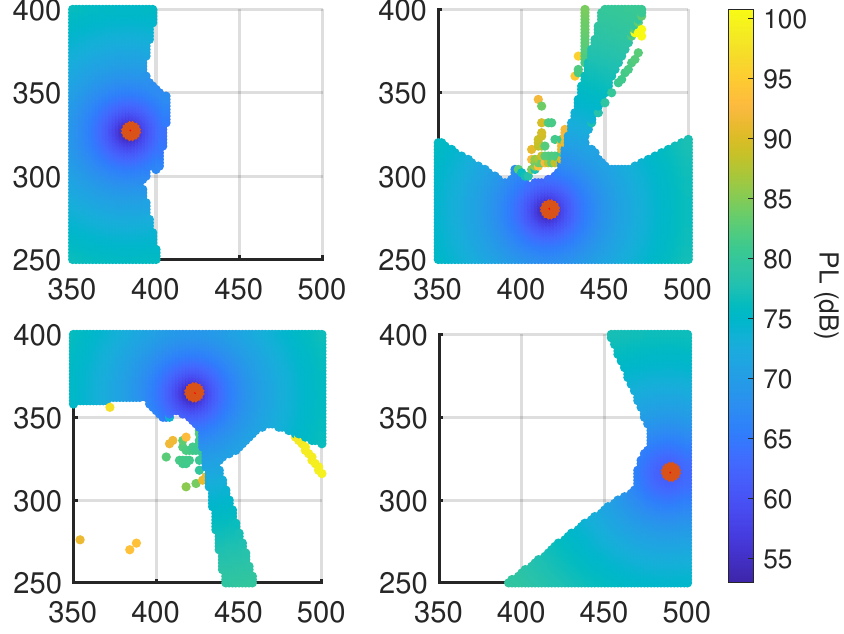}\label{fig:PLREC}}
\caption{\ac{pl} maps: (a) ground truth \ac{ckm}, obtained from ray tracer on the true map (Fig.~\ref{fig:MapGT}) and (b) ISAC \ac{pl} \ac{ckm} estimated on the ISAC-reconstructed map (\ref{fig:MapREC}), at the four  \acp{bs} (red dots).}
\label{fig:CKM}
\end{figure*}


In detail, likelihood \eqref{eq:likelihood} has been factored by leveraging the independence between the \ac{pl} and \ac{aoa} measurements at different \acp{bs}. Leveraging \eqref{eq:CLTPL} and \eqref{eq:CLTAoA}, it yields that both features are Gaussian distributed. Thus, considering for instance, the \ac{pl}, the likelihoods are 
\begin{equation}\begin{split}\label{eq:PLlikelihood}
  & p\Big(\hat{r}_n | r_n={r}_n(\bm{p})\Big)   =\\
    &=\begin{cases}
        P_{\emptyset, n}, \quad \,\, &\mbox{if } \hat{r}_n = {r}_n(\bm{p})=\emptyset \,,\\  
        0, \quad \quad &\mbox{if }\hat{r}_n \neq r_n(\bm{p}) = \emptyset\,\,,  \\
        0, \quad \quad &\mbox{if } r_n(\bm{p})\neq \hat{r}_n = \emptyset\,,  \\
                (1 - P_{\emptyset, n}) f (\hat{r}_n;{r}_n(\bm{p}),\hspace{-.3cm}& \sigma_{\mathrm{r},n}(\bm{p}) ),\mbox{otherwise} ,
    \end{cases}
\end{split}\end{equation} 
where $f(x; \mu ,\sigma)$ is the pdf of the Gaussian distribution with mean $\mu$ and standard deviation $\sigma$, and  $ P_{\emptyset,n}$ counts the fraction of how many positions in $\mathcal{P}_{\rm A}$ have signal obstructions for \ac{bs} $n$.
An expression analogous to \eqref{eq:PLlikelihood} is computed for $p\left(\hat{{\theta}}_n | {\theta}_n={\theta}_n(\bm{p})\right)$.

Finally, the authenticity is decided by the test function $   \hat{\mathcal{H}}_j = 
        \mathcal{H}_0$ if $\gamma\geq \xi$m and $\hat{\mathcal{H}}_j = \mathcal{H}_1$ if $\gamma < \xi$, where $\xi$ is a suitable threshold.
As customary, we evaluate the test performance by computing false alarm, i.e., the probability of labeling as fake the legitimate signal, $P_{\rm fa} = P[\hat{\mathcal{H}_1}|\mathcal{H}_0]$ and the missed detection, i.e., the probability of labeling as legitimate a signal transmitted by Trudy, $P_{\rm md} = P[\hat{\mathcal{H}_0}|\mathcal{H}_1]$.



\section{Numerical Results}\label{sec:NumResults}

\begin{table}
    \centering
    \caption{Default Simulation Parameters}
    \label{tab:simParameters}
    \begin{tabular}{ccc} \toprule
        \textbf{Parameter} & \textbf{Description} & \textbf{Value}  \\ \midrule
        $N$ & Number of \acp{bs}& $4$\\
        $N_{\rm r}$ & Number of \ac{bs} Rx antennas & $32\times32$\\
        $N_{\rm t}$ & Number of \ac{bs} Tx antennas & $1\times1$\\
        $N_{\rm p}$ & Number of pilots symbols & $10$\\
        $(P_{\mathrm T})_{\rm dBm}$ & \ac{ue} Tx power & $26\,$dBm\\
        $(G_{\mathrm{T}})_{\rm dBi}$ & \ac{ue} antenna gain & $2\,$dBi\\
        $(G_{\mathrm{R}})_{\rm dBi}$ & \ac{bs} single antenna element & $5\,$dBi\\
        $(\sigma^2_{{0}})_{\rm dBm}$ & Unscaled noise power \ac{bs}& $-55\,$dBm\\
        $f_{\rm A}$ & \ac{ue} carrier frequency& $3.5\,$GHz\\
        $f_{\rm I}$ & \ac{isac} carrier frequency& $26\,$GHz\\ 
        $B_{\rm I}$ & \ac{isac} bandwidth & $400\,$MHz\\ 
        $d$ & Antenna elements spacing & $\lambda/2$\\
        $W_{\rm M}$ & \ac{ckm} squares size & $1\,$m\\
        $W_{\rm A}$ & Alice location uncertainty& 
        $2 \, W_M$\\ \bottomrule
    \end{tabular}
    \vspace{-.5cm}
\end{table}

In this Section, we evaluate the proposed \ac{isac}-based \ac{ckm}-based \ac{pla} framework. In particular, we use the Sensiverse dataset~\cite{Jiajin-2023}, specifically developed for the evaluation of \ac{isac} systems, e.g.,~\cite{Song-2024}. Positions of the \acp{bs} and area size are shown in Fig.~\ref{fig:ISAC_reconstructed3D}, and simulation parameters are reported in Table~\ref{tab:simParameters}. We note in particular that sensing and channel estimation are performed at two different frequencies.

\begin{figure}
    \centering
    \includegraphics[width=0.9\linewidth,trim={3.4cm 8.5cm 4.4cm 9.0cm}, clip]{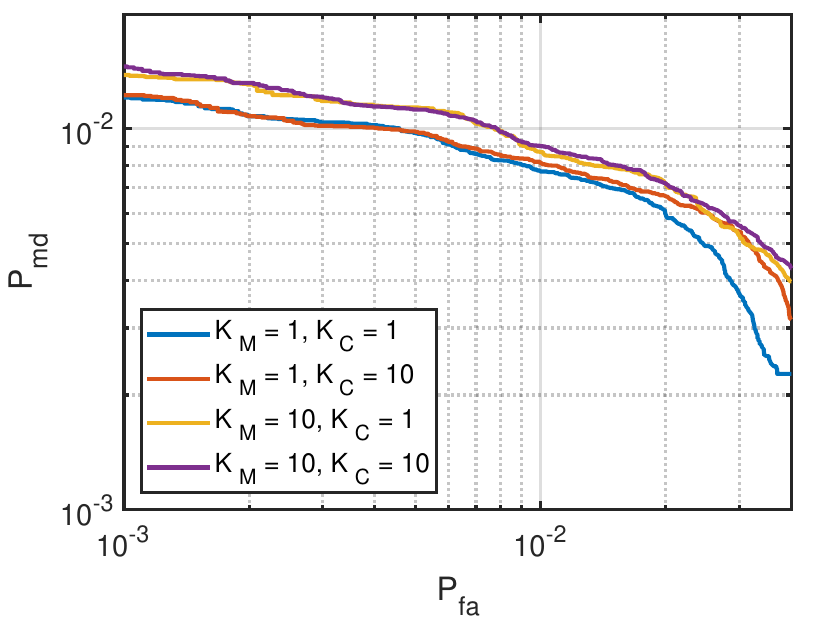}
    \caption{\ac{det} curve for different levels of noise $K_{\rm M}$ and $K_{\rm C}$, for $W_{\rm} = \SI{2}{\meter}$.}
    \label{fig:DET_Solution_comparison_1}
\end{figure}
\subsection{CKM Reconstruction Performance}

First, Fig.~\ref{fig:ISAC_reconstructed3D} compares the reconstructed environment with the ground-truth. The \ac{isac} system can effectively detect walls and obstacles facing the \ac{bs}. However, as expected, it encounters difficulties when reconstructing the surface and areas that are not directly visible from the \ac{bs}. Since the \acp{bs} are deployed around the environment rather than uniformly covering the entire scene, the system is not always able to recover the full shape of the structures. Nevertheless, the reconstructed environment still provides sufficient geometric information to capture the main characteristics of the propagation scenario.

\subsection{Authentication Performance}

Next, we assess the impact of environmental reconstruction on the quality of \ac{ckm}, i.e., verify whether ISAC can be used to generate reliable \acp{ckm}. To this end, Fig.~\ref{fig:CKM} compares in reference \ac{pl} \acp{ckm}, obtained from the ray tracer applied in the true environment, with the \ac{pl} \acp{ckm} generated from the reconstructed scene, at different \acp{bs}. The results show that the reconstructed \ac{ckm} struggles to accurately predict the channel in the regions between buildings, mainly due to imperfections in the underlying 3D environment reconstruction. However, outside these critical areas, the reconstructed maps provide a good approximation of the reference \ac{ckm}, preserving the main spatial variations of the channels throughout the environment.

Now, we evaluate the authentication performance of the proposed \ac{isac}-based \ac{pla} scheme. The authentication decision is made using the \ac{glrt} test described in Section \ref{sec:PropScheme}, which combines \ac{ckm} and the partial knowledge of Alice's position, e.g., obtained from the network, to decide whether the received signals are authentic.
The performance is evaluated in terms of \acf{det}, thus showing the missed detection probability $P_\mathrm{md}$ as a function of the false alarm probability $P_\mathrm{fa}$. The results are obtained through Monte Carlo simulations with $10^6$ independent trials.

Let the power of the \ac{awgn} noises for the \ac{ckm} reconstruction and \ac{pla} channel estimation be $\sigma_{\rm M} = K_{\rm M}\sigma_0$ and $\sigma_{\rm C} = K_{\rm C}\sigma_0$ respectively. Then Fig.~\ref{fig:DET_Solution_comparison_1} shows the \ac{det} for different pairs of $(K_{\rm M}, K_{\rm C})$, controlling the noise on the \ac{ckm} feature estimation and on PLA feature estimation, respectively. The $K_{\rm M}$ affecting the quality of the \ac{ckm} appears to have a more relevant impact than $K_{\rm C}$, thus suggesting that it is worthwhile to devote more effort (e.g., increase the number of pilots) during the ISAC-assisted \ac{ckm} estimation phase. On the other hand, once the \ac{ckm} are estimated, we can have a lightweight and reliable \ac{pla} check, e.g., even with a relatively low amount of transmitted pilot signals.

\begin{figure}
    \centering
    \includegraphics[width=0.9\linewidth,trim={3.4cm 8.5cm 4.4cm 9.0cm}, clip]{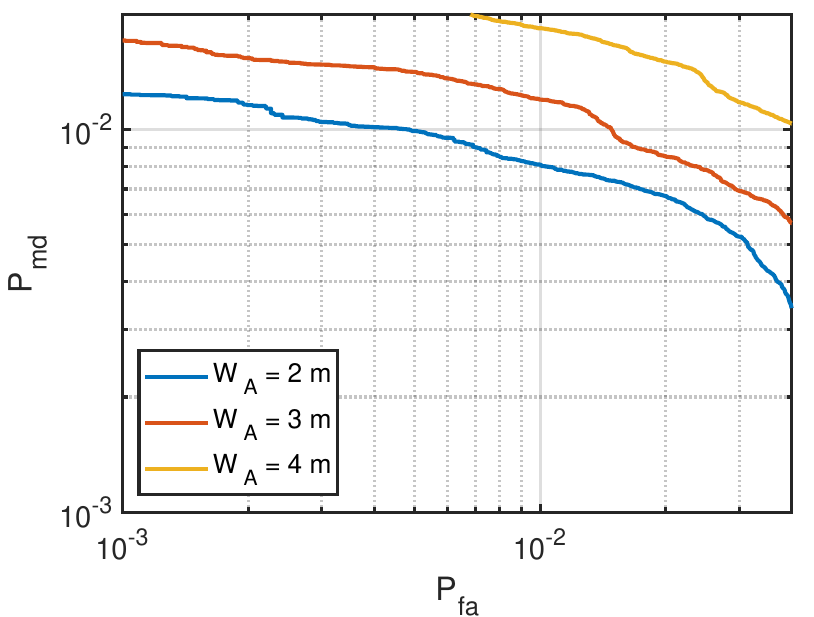}
    \caption{\ac{det} curve for different $W_{\rm A}$ values. $K_{\rm M} = 1$ and $K_{\rm C} = 1$.}
    \label{fig:DET_Solution_comparison_2}
\end{figure}

Fig.~\ref{fig:DET_Solution_comparison_2} reports the \ac{det} for different values of $W_{\rm{A}}$ with $K_{\rm C} = K_{\rm M} = \SI{0}{\decibel} $, to evaluate the impact of accuracy on the Alice position on the overall \ac{pla} scheme. 
As expected, higher $W_{\rm A}$ values, and thus higher uncertainty, lead to worse results. Still, the decrease is controlled and can be countered by improving the quality of the estimated features, e.g., increasing $K_{\rm M}$.  
It is worth remarking that the perfect knowledge case was tested (i.e., with $W_{\rm A} = \SI{1}{\meter}$), but has been omitted in the plot, as it achieved performance much lower than working point, i.e., $P_{\rm MD}\ll 10^{-3}$ for the considered range of $P_{\rm FA}$ values.

\balance
\section{Conclusions}\label{sec:Concl}

In this paper, we proposed an \ac{isac}-enabled framework that reconstructs the physical environment to build a data-driven \ac{ckm} via ray tracing for \ac{pla} applications. By comparing real-time estimated channels with location-dependent \ac{ckm} entries, the system effectively authenticates a legitimate transmitter and detects spoofing devices from alternative locations. Furthermore, we analyzed how \ac{isac} reconstruction errors and receiver-side channel estimation noise impact system performance. Validated against a literature-sourced \ac{csi} dataset, numerical results demonstrate that our framework maintains robust security even under non-ideal conditions, achieving false alarm and missed detection probabilities below $10^{-2}$. This confirms the viability of combining environmental sensing with channel-aware positioning to secure future wireless networks.

\bibliography{IEEEabrv,biblio.bib}
\bibliographystyle{IEEEtran}

\end{document}